\documentclass[twocolumn]{aa}
\usepackage{graphicx}
\begin{document}
\title{Why  Hot Horizontal Branch Stars Can Appear Redder Than Red Giants
\thanks{Based on  observations with the ESO/MPI  2.2m
telescope, located at La Silla Observatory (Chile) and on observations 
with the NASA/ESA {\it Hubble Space Telescope}.}} 

\author{Y. Momany\inst{1}, S. Cassisi\inst{2}, G. Piotto\inst{1}, L.R. 
Bedin\inst{1},  S. Ortolani\inst{1}\\ F. Castelli\inst{3,4}
\and A. Recio-Blanco\inst{1}
}

\offprints{Y. Momany}

\institute {Dipartimento di Astronomia, Universit\`a di  Padova,   
Vicolo  dell'Osservatorio  2, I-35122     Padova, Italy\\
\email{momany,piotto,bedin,ortolani,recio@pd.astro.it }
\and 
INAF - Osservatorio Astronomico di Collurania, Via M. Maggini,   
64100 Teramo,  Italy\\ 
\email{cassisi@te.astro.it}
\and  Istituto di Astrofisica Spaziale e Fisica Cosmica, CNR, Via del
Fosso del Cavaliere, 00133, Roma, Italy  
\and  Osservatorio Astronomico di Trieste, via Tiepolo 11, 34131 
Trieste, Italy \\ 
\email{castelli@ts.astro.it}
}

\date{Received March .., 2003; accepted March .., 2003}

\abstract{

In this  paper we report  on a  curious feature  in  the $V$,  ($U-B$)
color-magnitude diagrams of   globular clusters.  In our database,  we
find that  a  considerable fraction of  blue  horizontal branch stars,
hotter than the instability strip and  cooler than the Grundahl et al.
(1999) jump  (i.e.,  $6000 \le T_{\rm  eff}$(K)$\le 10000$),
have ($U-B$) colors redder than their red giant progenitors.  This red
incursion is not expected on theoretical grounds, as horizontal branch
stars (whose  convective regions are less extended  than  in red giant
structures) should not ``appear'' cooler than a red giant.
Analyzing data from different     telescopes  we show that:  1)    the
horizontal branch red incursion is strongly dependent  on the shape of
the adopted $U$ filter and  to a lesser  extent, on the $B$ filter; 2)
the photometry done with $U$ filters which do not encompass the Balmer
jump, do not  show  the blue  horizontal branch red  incursion; 3) the
occurrence of this feature  is also due to  the peculiar dependence of
the $U$ and  $B$ magnitudes on  stars effective temperature,  gravity,
and  metallicity;  4) theoretical  tracks   can reproduce the observed
horizontal  branch morphology,   provided  that the  appropriate (i.e.
exactly    responding to   the    filters  effectively  used in    the
observations)  transmission curve  efficiencies are  used for deriving
color-$T_{\rm  eff}$   transformations; 5) the   red  incursion extent
depends on metallicity.

}

\authorrunning{Momany et al.}

\titlerunning{The blue horizontal branch red incursion}

\maketitle

\section{Introduction}

Color-magnitude diagrams  (CMD)  of resolved  stellar populations are
the fundamental tool  for testing predictions of stellar  evolutionary
theory and the  capability of stellar models  to finely reproduce  the
observations.  Photometric studies    are  no longer  limited to   the
classical $BVI$ bands,  as high quality  near-infrared and ultraviolet
CMDs are becoming more and more available.
Ultraviolet (UV) bands certainly cover the ideal wavelength region for
the study of hot stellar populations.  Indeed, UV observations help to
remove  the degeneracy  between  color and   temperature for hot  blue
horizontal branch (HB) stars, allowing  a more reliable measurement of
atmospheric parameters (in particular temperature estimates).

\begin{figure*}[t]
\centering
\includegraphics[width=18cm,height=20cm]{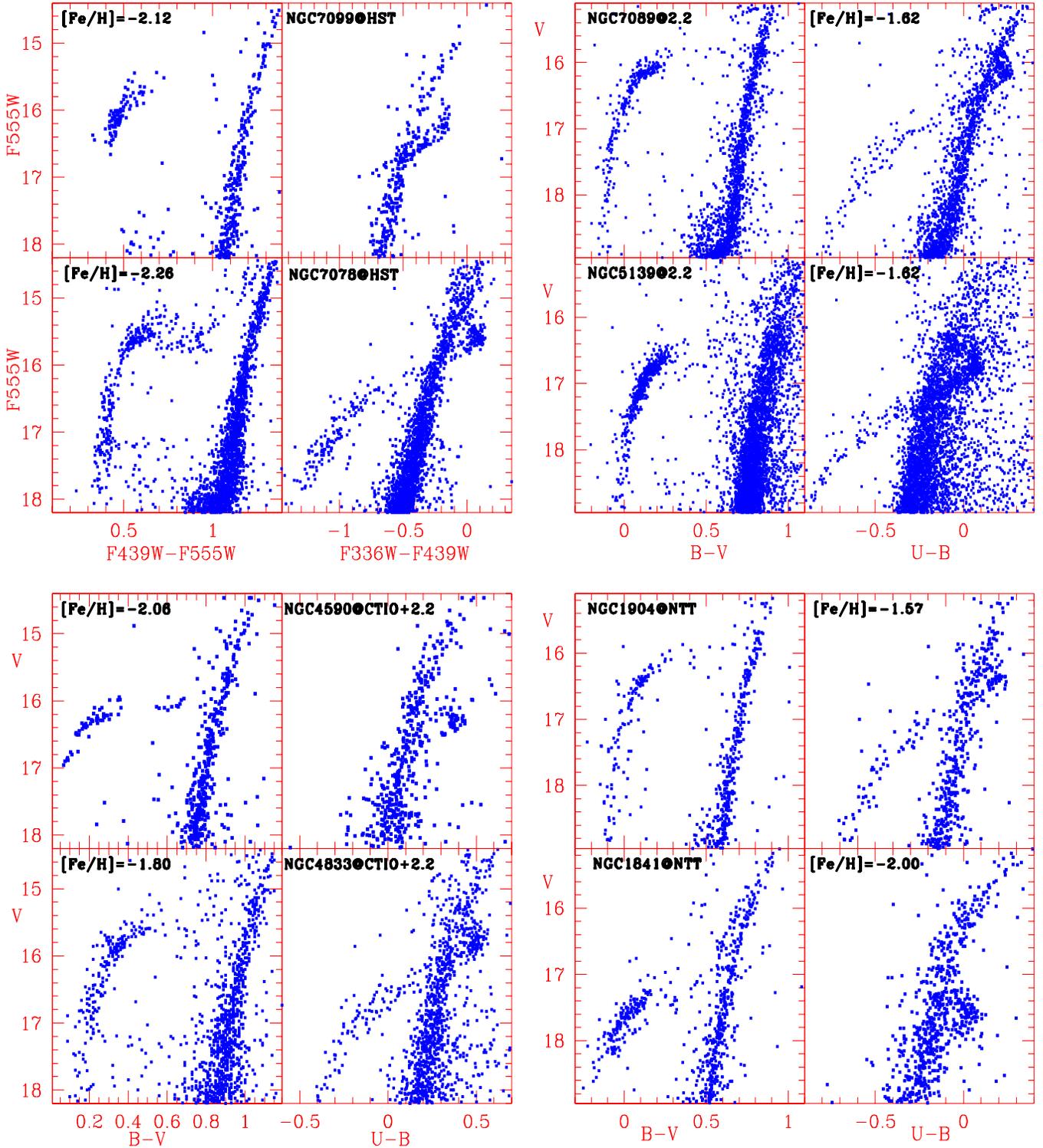}
\caption{$V$, ($B-V$) and $V$, ($U-B$) diagrams of a BHB cluster sample
 collected at 4 different telescopes.  The diagrams in each panel were
 shifted for demonstration  purposes.  The reported metallicities  are
 from          the       Harris          on-line-catalog:         {\sf
 http://physun.physics.mcmaster.ca/$\sim$harris/mwgc.dat}
 (\cite{harris96})  as updated  on  February 2003, except for  NGC1841
 (Alcaino et al.  1996).}
\label{f_all6}
\end{figure*}

Recently, the blue  horizontal  branch  (BHB)  population has  been  the
subject of specific  UV studies,  and  some interesting features  were
revealed  only     in  ultraviolet    CMDs:     jumps   (Grundahl   et
al. \cite{grun99},  G99,  Momany et al.   \cite{moma02}), and peculiar
horizontal branch   extensions,   like the blue-hook (D'Cruz    et al.
\cite{cruz00}, Brown et   al.  \cite{brow01}).  The ultimate   goal of
these  studies  is a  better understanding of  the physical mechanisms
governing core-helium burning stars, and driving the horizontal branch
morphology.  To this end, our group has undertaken a long-term project
to obtain  wide-field multi-wavelength  data of BHB  globular clusters
(GCs). Mainly, we used  the WFPC2 on board of  HST to map the  central
core of our target clusters  (Piotto et al.  \cite{piotto02}), and the
wide  field imager (WFI)  at the ESO/MPI 2.2m  for the coverage of the
outer  regions.   Unfortunately,  observations  carried   out  through
non-standard  photometric  systems  sometimes   generate puzzling  and
apparently un-physical properties in the observed CMDs.

In this paper  we  report  on a  curious  feature in  the observed  HB
morphology in the UV color-magnitude diagrams,  and an explanation for
it.  Basically, in some of the $V$, ($U-B$) diagrams of our database, we
find that  the  blue HB stars hotter  than  the  instability strip and
cooler than  the G99 $u$--jump  ($-0.1\le  B-V \le0.2$) have
$U-B$ colors that  place them on the right  hand side of the red giant
branch (RGB).

According  to  stellar evolutionary theory  blue HB  stars  should not
``appear'' cooler  than their RGB  progenitor.  Indeed, HB  stars have
hotter effective temperature than RGB stars with the same metallicity.
In a  color-magnitude diagram, we  do not expect  any red incursion of
BHB stars to  the cooler side of the  RGB (which almost corresponds to
the Hayashi  track): for stars  in hydrostatic  equilibrium and having
convection, the Hayashi  line separates a  {\it  permitted} region (on
its left)  from a {\it  forbidden} one  (on its right).   Despite this
theoretical  constraint, we will show that  the  unexpected finding of
the BHB  red incursion is  present in at least 6  $V$, ($U-B$) CMDs of
BHB clusters we have observed. We will also show that this peculiarity
is not present  in all  the CMDs  from  broad band  photometry of  BHB
clusters, as the effect depends on the metallicity.

As  discussed in Section~3,  it  turns out  that the apparent  anomaly
rises  when  ``non-standard'' $U$  filters are   used.  We  decided to
publish these color-magnitude diagrams, and discuss  the origin of the
apparently  un-physical HB morphology,  in order to call the attention
on the  importance  of using appropriate  filter  sets  when acquiring
photometric data.
Alternatively, and for a proper interpretation  of the empirical data,
this  paper shows  the  importance of   using exactly the  photometric
passbands   of   the  observational   system  when    transforming the
theoretical tracks to the observational plane.

\section{Observations and Data Reductions}
\label{s_obs}

We  present data collected with  the Hubble Space Telescope (HST), and
the  ESO/MPI 2.2m telescope.  $UBV$  data of  NGC7089 and NGC5139 were
taken on June  2002 with the  Wide-Field-Imager  (WFI) at the  ESO/MPI
2.2m, employing  the  ``new'' $U$ filter  (Germany  \cite{germ02}).  A
second $UBV$ data  set of NGC7078 and NGC7099  images was obtained  on
July 2000, employing the ``old'' $U$ filter at the same telescope.

Basic reduction of the CCD mosaic was performed using the IRAF package
MSCRED   (Valdes 1998),  while    stellar photometry  was done   using
DAOPHOT/ALLFRAME  (Stetson 1994).  Calibration  to the  standard $UBV$
system was   obtained  using a set  of  standard   stars from  Landolt
(1992).  HST data of NGC7099, NGC7078,  and NGC6205 were  taken in the
F$336$W, F$439$W, and F$555$W bands. For these clusters we present only
instrumental diagrams. Besides the HST and 2.2m data, we present $UBV$
published data of  NGC1904   by Kravtsov et al.  (\cite{krav97}),  and
NGC1841 by  Alcaino  et al.   (\cite{alca96}),  both collected  at the
ESO/NTT telescope. 
Lastly, we  present composite $UBV$  catalogs of  NGC4833 and NGC4590.
The $BV$ photometry of these clusters were taken from Melbourne et al.
(\cite{mel00}) and  Walker (\cite{walker94}), while the $U$ photometry
were  obtained employing  the  WFI  ``new''  $U$ filter   at  the 2.2m
telescope.

\section{The Blue Horizontal Branch Red Incursion}

Figure~\ref{f_all6} shows a zoom around the HB in the $V$, ($B-V$) and
$V$, ($U-B$)  diagrams  of the 8  GCs, obtained  with the 4  different
telescopes (HST, NTT, CTIO and 2.2m ESO telescope).  The clusters span
a range  in metallicity of $\sim  0.7$ dex, from the metal-poor regime
to intermediate metallicity.   For  all clusters  in the  sample,  the
presence of  the HB elbow  and its incursion  through the RGB is quite
evident. Due to the larger sample of HB stars, it is really impressive
in  NGC7089, NGC7078,  and  NGC5139\footnote{$I$, ($U-B$)  diagrams of
these clusters show even a greater extent of the red incursion}.  Note
how the blue HB of NGC7099 lies almost entirely on the red side of the
RGB in the $V$, ($U-B$) diagram.

\begin{figure}[h]
\includegraphics[width=9cm]{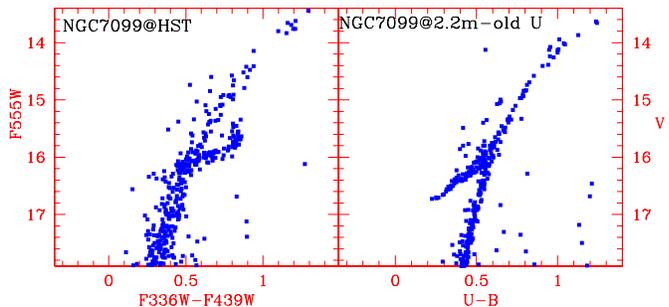}
\caption{Color-magnitude diagrams of NGC7099 showing completely 
different HB morphologies as a function of the employed filters.}
\label{f_7099}
\end{figure}
\begin{figure*}[t]
\centering
\includegraphics[width=16cm,height=16cm]{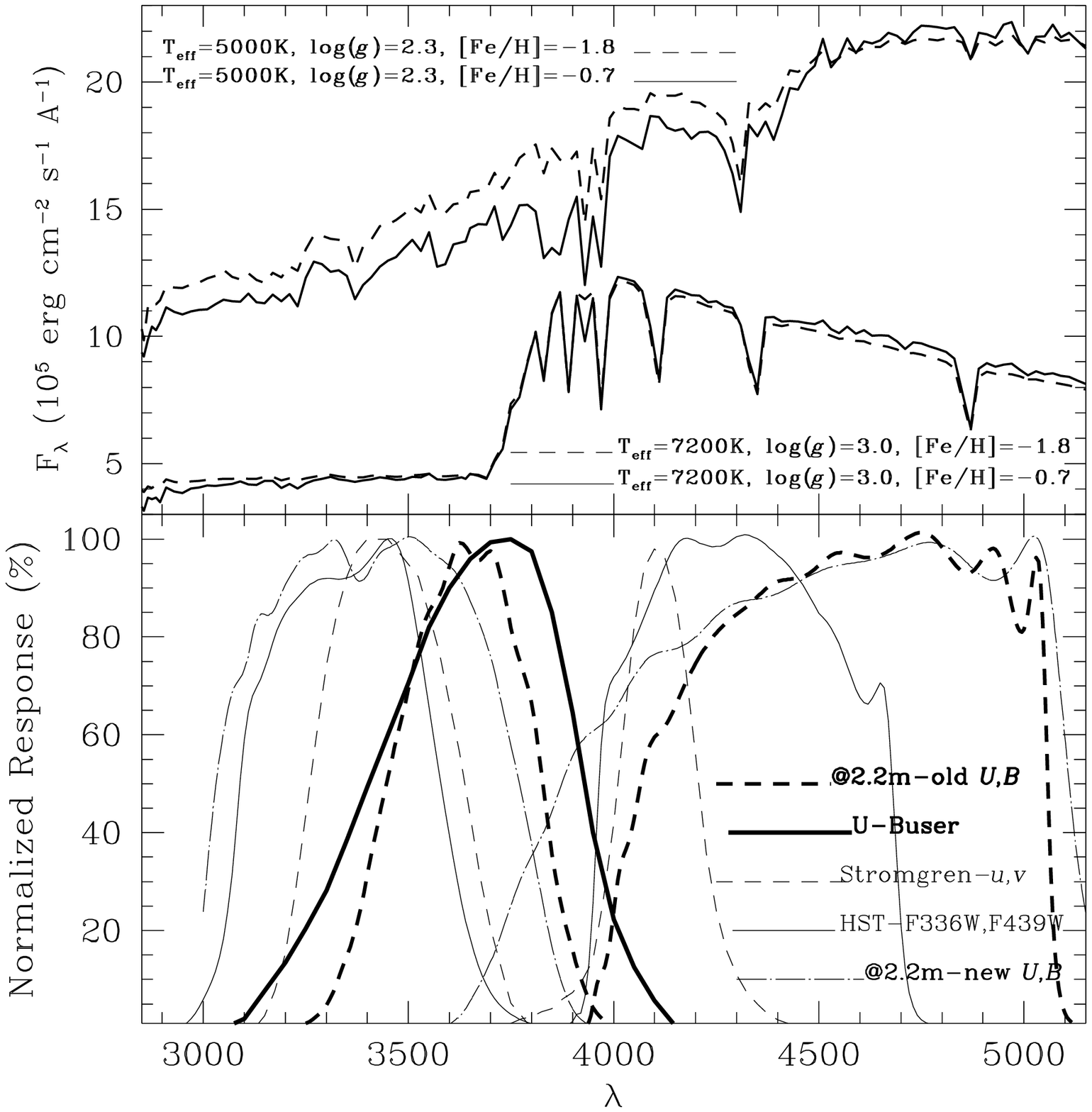}
\caption{Upper panel shows a comparison between different stellar 
spectra for two different choices of the star metallicity, gravity and
effective  temperatures, representative of  BHB  stars (lower spectra)
and RGB structures (upper spectra) with the same bolometric luminosity
of HB  objects.  The  ``RGB''  spectra  has been shifted  upwards  for
demonstration purposes.  Lower panel shows the normalized transmission
curves of different $UB$ filters.}
\label{f_z1.8}
\end{figure*}

Figure~\ref{f_7099} shows some of  the conflicting results we faced in
the   course  of our  investigation.     The $V$,  ($U-B$) diagrams of
NGC7099, from   data collected at     two different telescopes,   show
completely  different  HB morphologies.  In  the   case of the diagram
obtained with HST, the BHB incursion  is clearly visible, while in the
other diagram (based on  data collected with  the 2.2m  telescope) the
BHB does not cross the cluster RGB at  all.  These differences are not
to be  attributed to the  limit  of the UV  region  reachable from the
ground   (at   variance  with   the   HST   space  observations),   as
Fig.~\ref{f_all6} shows two CMDs (for  NGC7089 and NGC5139), from data
obtained at the ESO/MPI 2.2m  telescope, clearly disclosing the HB red
incursion.  Neither the effect can be attributed to red leakage in the
employed $U$  filters  (the WFPC2  F336W and  WFI  new-$U$ filters are
known to be  affected by red-leaks).   Red leakage is affecting mainly
RGB stars,  causing them to have  smaller than expected $U$ magnitudes
and therefore bluer $U-B$ colors. Indeed, considering the F336W filter
(the most affected by red leakage among our filters) we calculate that
for  an RGB star at  the level of the  HB, with [Fe/H]$=-1.8$, the red
leakage  in the F336W band  causes a blue-ward  shift  of only $0.012$
magnitude while the effect we discuss in this paper is of the order of
few tenths of a magnitude.

The apparent contradictory results shown  in Fig.~\ref{f_7099} lead us
to  search for differences  among  the filters adopted  to acquire the
different data sets, in particular at the ESO/MPI  2.2m telescope.  We
found that the 2.2m data showing the BHB incursion were obtained using
the  two ``new'' $U$  and  $B$ filters (Germany \cite{germ02}),  while
data not showing the incursion were  obtained with the ``old'' $U$ and
$B$ filters.

Prompted by this result,  we have analyzed the normalized transmission
curves of  a  variety of  $U$ and $B$  filters (Fig.~\ref{f_z1.8}). We
found that the  ``new'' $U$ filter at  the 2.2m,  as  well as  the HST
F336W, and the Str\"omgren $u$ filters,  have a transmission curve with
a peak around $\lambda\simeq 3400 \AA$. On the other hand, the ``old''
$U$ filter      at   the  2.2m,   and    the   $U-$Buser   band (Buser
\cite{buser78}) of the  synthetic colors  (Castelli \cite{cast99}) that
were originally used for  the transformations from the  theoretical to
the observational plane,   have  a transmission curve  peaking  around
$\lambda   \simeq 3650 \AA$.  It  is  worth mentioning that the latter
group  of filters have  transmission   curves that  are closer to  the
standard $U$ Johnson (although this does not necessarily mean that the
observational  band,  which  is  a convolution   of the  filter,  CCD,
telescope  optics, and  atmosphere transmission curves  is exactly the
standard $U$ Johnson (\cite{john55}) band).
Interestingly enough, the  clusters observed with  the latter group of
filters do not  show   the BHB  incursion, whereas  clusters  observed
through  filters whose transmission  peak at $\lambda \simeq 3400 \AA$
do.    Also the $B$ filter   can be important.   For example, clusters
observed with the new 2.2m $U$ filter and the old 2.2m $B$ filter (see
Fig.~\ref{f_z1.8}) do {\em not} show the incursion. Clearly, it is the
combination of non-standard $U$   and  $B$ passband that creates   the
observed anomaly.

In Fig.~\ref{f_z1.8} we show  the normalized transmission curve of the
$UB$ filters  used at the 2.2m,  the HST F336W  and F439W filters, and
the Str\"omgren $u$ and $v$ filters.   For two different metallicities
([Fe/H]$=-1.8$ and $-0.7$), we also show synthetic spectra (Castelli
\& Kurucz \cite{cast03}) for BHB stars, and for RGB stars with surface
gravity and  effective temperature  appropriate  of red giants  at the
level of the HB luminosity.
When comparing    a BHB  spectrum   with  the  transmission  curves of
different $U$ and $B$ filters, it is  clear that the occurrence of the
BHB red incursion is related  to the Balmer discontinuity: $U$ filters
peaking around $\lambda  \simeq 3400 \AA$ do  not encompass the Balmer
jump. These filters collect less $U$ flux with  respect to $U$ filters
peaking at $\lambda \simeq 3650 \AA$. Consequently, the $U$ magnitudes
of BHB  stars are fainter and therefore  their ($U-B$) colors  will be
redder.  On the  other hand, RGB  stars have a shallower Balmer  jump,
hence these are less affected by differences in $U$ filters.
Also important is the
broadness  of the  new-$B$   filter at the 2.2m   with  respect to the
old-$B$ filter.   The blue side of  the new-$B$ encompasses the Balmer
jump, and is the cause of the anomalous morphology  of the HB when the
corresponding  $B$ magnitudes are  combined with new-$U$ filter at the
2.2m.

\begin{figure}
\includegraphics[width=9cm]{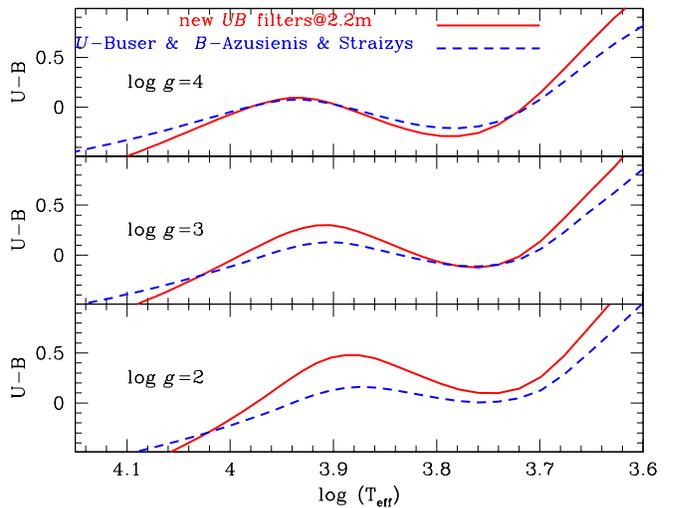}
\caption{Computed $T_{\rm eff}-$($U-B$) relations for [M/H]$=-2.0$, 
different log~$g$, and different $UB$ filters.  {\it Dashed} lines are
the  relations  obtained   using  the  $U$-Buser (\cite{buser78})  and
$B$-A{\v z}usienis  \&   Strai{\v z}ys  (\cite{azus69})  filters, {\it
continuous} lines are those obtained using the new $UB$ filters at the
2.2m telescope.}
\label{f_cast}
\end{figure}

A  deeper understanding of the BHB  red incursion can be achieved when
considering the dependence of the  ($U-B$) color on stellar parameters
such   as   surface   gravity, effective    temperature,  and chemical
composition.
The $UBV$ synthetic photometry from the ATLAS9 models (Castelli
\cite{cast99}, Fig.~11) shows   that, for  a  given gravity,   the
($U-B$)    index does  not    monotonically  increase with  decreasing
temperature    (see    also   Fig.~\ref{f_cast}).   Instead,   between
$\sim6750-5500$K,  the ($U-B$) index   shows a minimum which increases
with decreasing metallicity,  and is almost independent from  gravity.
Moving towards higher  temperatures  ($\ge7000$K), the   ($U-B$) index
increases from the quoted minimum  and reaches a maximum value,  whose
intensity  depends on gravity, but  not on metallicity.   The lower is
the gravity, the higher is  the maximum. Moreover, the ($U-B$) maximum
moves  from   $\sim7500$K for    log~$g=2.0$   to about  $9000$K   for
log~$g=5.0$.
This means that in a  metal-poor cluster, the  ($U-B$) index of a star
with $T_{\rm eff}\approx8000$K and  log~$g=3.0$ (i.e. an HB  star) may
be close to the maximum value and  therefore exceeds the ($U-B$) index
of a RGB star with $T_{\rm eff}\approx5500-6000$K.

This  anomalous  ($U-B$)  behavior    is not  encountered    when  the
``standard''  $U$-Buser  filter  or the old-$U$    filter at the  2.2m
telescope are used, simply because the  difference between the maximum
and the minimum in the $T_{\rm eff}-$($U-B$) curve is not large enough
to produce the red  incursion.    On the  other hand,  the  difference
between the minimum  and maximum ($U-B$)  index becomes larger for $U$
filters with transmission curves  shifted towards shorter wavelengths,
to the  point of producing the  red incursion for  the F336W filter on
HST,  Str\"omgren $u$  filter,  and  the new   $U$ filter  at the 2.2m
telescope.

%
To better explain the red incursion observed for low gravity stars, we
replaced in the synthetic photometry (Castelli, \cite{cast99}) the $U$
filter  from Buser (\cite{buser78})  and   the  $B$ filter from   A{\v
z}usienis \& Strai{\v  z}ys (\cite{azus69}) with the  new $U$  and $B$
filters of the 2.2m photometric system.
Figure~\ref{f_cast} shows that,  for   typical gravities of   HB stars
(log~$g=3.0$),  there     is a  large     difference  in  the  $T_{\rm
eff}-$($U-B$) curves for the two sets of  filters; particularly at the
($U-B$) maximum, where the difference is $\sim0.2$ mag.
Figure~\ref{f_cast} also shows  that while the  difference between the
($U-B$)   index of   an  HB   star  (log~$g=3.0$)  and   an RGB   star
(log~$g=2.0$) for the classical  filters is approximately the same for
all the temperatures, for the new $UB$ filters at the 2.2m the ($U-B$)
color of an HB star is {\it redder} than the ($U-B$) of an RGB star.

The effect we have just  discussed is not new.  It  has been known  as
the  ``ultraviolet deficiency''  of  cool,  BHB  stars in the  ($U-B$,
$B-V$) two-color   diagram (see   Markov  et al.   \cite{mark01}   and
references therein).  Analyzing data for 6 GCs, Markov et al. compared
the position of BHB stars with respect to  i) a Population~I sequence;
and ii)  a   zero-age horizontal branch  (ZAHB)   two-color  line, and
quantified the  $\Delta  (U-B)$  deficiency.    In analogy   with  our
results, they also found that this observable is not common among {\em
all} BHB clusters.   Markov   et  al.   argued  that  there  exists  a
connection  between  this   observable and    cluster parameters.   In
particular, they showed how a satisfactory overlapping of the observed
and theoretical sequences is achieved when using low-gravity (log~$g
\le 2.0$) atmospheric models.

As expected from the above discussion,  using a $U$ filter-set showing
the BHB red incursion  (e.g.  the F336W filter on  HST or  the ``new''
$U$ filter  at the 2.2m  telescope) not all BHB  clusters show the red
incursion.
It  is clearly present in a  metal-poor  cluster like NGC7078, whereas
completely  missing  in   an intermediate  metallicity   cluster  like
NGC6205 (see also Fig.~\ref{f_isoch}).
Due  to the small sample of   clusters, we can neither observationally
quantify the dependence of the effect on the cluster metal content nor
precisely  estimate the cutoff  metallicity (if  any)  beyond which no
incursion is observed.
The fact that the   HB of NGC1904  ([Fe/H]$\sim-1.6$,  $E_{B-V}=0.01$)
shows   a    red    incursion,   while    NGC6205   ([Fe/H]$\sim-1.5$,
$E_{B-V}=0.02$) does not  show any incursion  could suggest that  this
effect  might be present mainly in   clusters with metallicities lower
than [Fe/H]$\approx -1.5$.
Inspection    of the HST F555W,    (F336W$-$F439W)  diagram of NGC6388
([Fe/H]$\simeq-0.60$, showing no indication of the incursion) suggests
that the effect is not present in the CMDs  from broad band photometry
of high metallicity BHB clusters.
In  view  of  the  strong  dependence  of  the  effect  on  the filter
transmission curve, only  a photometrically  homogeneous observational
data  set,  covering a large    metallicity  range will  allow us   to
characterize the phenomenon,  and investigate whether this feature can
be used  to  constrain  other cluster parameters,   like  metallicity,
distance, and reddening.

The suggestion for a metallicity threshold of the incursion relies on
a set of images collected with filters which are supposed to reproduce
the standard $UBV$ Johnson bands at 3 widely used telescopes: HST, NTT
and the 2.2m.
It must be   noted,  as pointed out   by  the referee, that  the   red
incursion effect   is  well visible  also  in other  bands,  like  the
Str\"omgren  intermediate ones. However,  the metallicity threshold is
not present in all the photometric systems.
Indeed, theoretical isochrones and ZAHBs  (from the Cassisi \& Salaris
(\cite{cassisi97}) library)  show  the  incursion  in  the Str\"omgren
M$_V$,  ($u-v$)   plane    even  at   metallicities   of    $Z=0.006$,
representative  of metal rich GCs.  
%

\begin{figure}
\includegraphics[width=9cm]{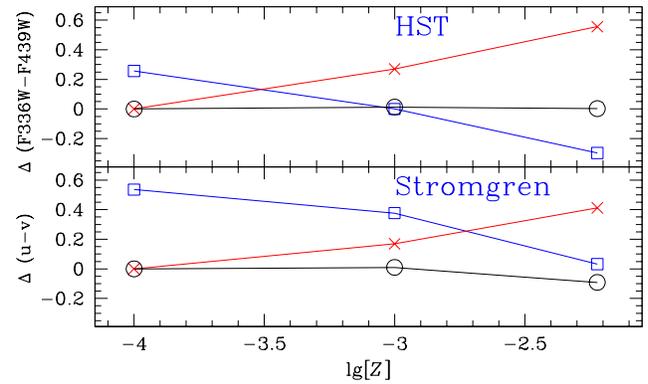}
\caption{RGB and HB color differences, at   the  level  of the    
reddest HB  incursion,  estimated  for three  isochrones  ($Z=0.0001$,
$0.001$,    and $0.006$)  in  the   HST   flight  and the  Str\"omgren
photometric systems.  Open squares measure the extent of the incursion
($\Delta$ color   between the HB  and  the RGB), crosses  mark the RGB
color  differences of   the  three isochrones    with respect  to  the
$Z=0.0001$ one,  and open circles mark  the HB color  differences with
respect to the $Z=0.0001$ isochrone.}
\label{f_strom}
\end{figure}

%
Figure~\ref{f_strom} helps to understand the metallicity effect of the
BHB red incursion.  For the HST flight and the Str\"omgren photometric
systems we  have plotted the  following  three color differences as  a
function of metallicity:
(1) $\Delta(HB-RGB)$  measures the extent of  the incursion, i.e.  the
difference in color between the reddest HB point  and the RGB color at
the level of the reddest HB point;
(2) $\Delta(HB-HB_{Z=0.0001})$ is the difference  in color between the
reddest HB point for the quoted metallicities and that for $Z=0.0001$;
and  lastly, 
(3) $\Delta(RGB-RGB_{Z=0.0001})$  is  the difference  between  the RGB
colors (at the level of the  reddest HB point  for the three indicated
metallicities) and that for $Z=0.0001$.
%
There   are    two  important  conclusions that    can   be drawn from
Fig.~\ref{f_strom}.   First of  all, we confirm   that,  while the red
incursion of  the  HB is  present at   least  up to  metallicities  of
$Z=0.006$ in   the Str\"omgren photometric  system, the  phenomenon is
expected to  disappear for clusters more  metal rich than $Z=0.001$ in
the  HST   flight system,  where   the  corresponding $\Delta(RGB-HB)$
becomes negative.
The second important evidence  disclosed by Fig.~\ref{f_strom} is that
the extension  of the red  incursion (and  its disappearance  for more
metal rich clusters in  the broad band CMDs) is  due to the  fact that
the RGB becomes redder and redder at increasing metallicity, while the
color of the reddest  HB point remains more less  the same  at varying
the cluster metallicity.

In Fig.~\ref{f_isoch}, we also  compare the observed  WFPC2/HST F555W,
(F336W$-$F439W)  CMDs of NGC7078 and  NGC6205 with the isochrones from
Cassisi \& Salaris (\cite{cassisi97})  transformed into the HST flight
system.  The data have not been calibrated, but the calibration to the
HST flight system  is simply a zero  point  translation, and does  not
affect the overall morphology of  the CMD.  It  is worth noticing that
the models reproduce very well the BHB ``elbow'' morphology, and, most
importantly, reproduce the BHB red incursion when present.
We have  also  compared the models  with   the CMDs from  ground-based
photometry for  those clusters showing the BHB  red incursion. In this
case the models are not able to reproduce such a feature at all.  This
is a consequence  of the fact that  the $U$-filter used to derive  the
color-$T_{\rm eff}$  relations  (adopted for  transforming the  models
from the theoretical to the  empirical plane) is the $U$-Buser filter,
clearly inadequate   to fit data  obtained   with ``non-standard'' $U$
filters (cf Fig.~\ref{f_z1.8}).

\begin{figure}
\includegraphics[width=9cm]{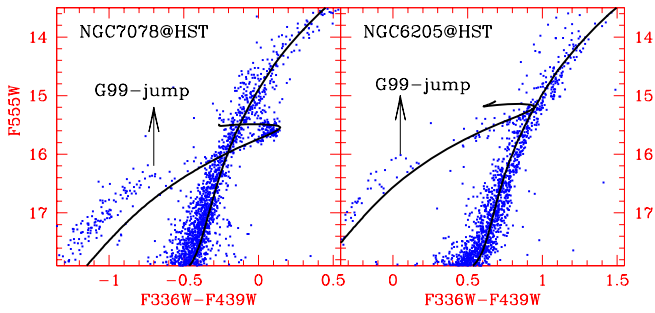}
\caption{HST F555W, (F336W$-$F439W) diagrams of NGC7078 
and  NGC6205, along with  theoretical isochrones and ZAHBs (Cassisi \&
Salaris  \cite{cassisi97}) with  $Z=0.0001$  (appropriate for NGC7078)
and $Z=0.001$ (appropriate for NGC6205).}
\label{f_isoch}
\end{figure}

This  occurrence   strongly  reinforces   the   need for   appropriate
color-$T_{\rm eff}$ relations, based    on  exactly the same    filter
passband used  in the observations, in  order to realize  a meaningful
comparison  between  theory and   observations.  Most importantly, the
anomalous  CMD morphology (produced  by $U$ filters whose transmission
is  far from the  standard  ones) calls  attention to another problem,
often ignored by those in charge of the development and maintenance of
imaging instruments: filter passbands must be as  close as possible to
the  bands  they pretend to   reproduce.  This simple rule  makes  the
calibration to  the  standard photometric   system more accurate,  and
avoids   the production of    apparently  un-physical features in  the
observed CMDs.

\section{Summary and final remarks}

In the present work, we show the existence of a curious feature in the
$V$, ($U-B$) CMDs of some  GCs.  Quite surprisingly, in some  observed
CMDs, hot  HB  stars ($6000\le{T_{\rm eff}}$(K)  $\le10000$)
become redder than  the much  cooler  RGB.  An investigation of   this
effect has shown the following:

\begin{itemize}

\item the size and the presence itself of this feature
is strongly dependent  on the passband of the  adopted $U$  filter. In
particular, $U$ filters which do not  encompass the Balmer jump break,
do not show the BHB red incursion;

\item also the $B$-filter passband concur to cause the anomaly;

\item the occurrence of this feature is due to the peculiar dependence   
of the $U$ and   $B$ filters on  star effective temperature, gravity, 
and metallicity;

\item clusters more metal-rich than [Fe/H]$\simeq-1.5$ do not show 
any BHB red incursion in the broad band Johnson and HST flight system,
while there  is     no metallicity threshold   in   Str\"omgren $uvby$
photometry for metallicities lower than $0.006$;

\item evolutionary models  using the correct filter passband, when 
transformed to the observational  plane, finely reproduce the observed
HB morphology.

\end{itemize}

We report this anomalous HB morphology to  point out the importance of
having  observational  photometric systems which  accurately reproduce
the  standard ones.  Unfortunately, the   fact that the anomalous CMDs
shown in this paper have  been obtained from  data acquired at some of
the most widely  used  instruments for astronomical imaging  shows how
this problem is yet to be addressed.

\begin{acknowledgements}
We  warmly thank the referee,  Frank Grundahl, for the useful comments
which helped to  improve the paper.  In  particular, we thank him  for
pointing out the  presence of the effect described  in this paper also
in the CMDs from Str\"omgren photometry.  We also thank G. Bono and V.
Castellani  for  useful discussions   on  this   matter.  Y.M   thanks
S. Lucatello for useful discussions  on stellar spectra. L.R.B., Y.M.,
S.O.,   G.P.,  and  A.R.B.   acknowledge   the   support  of the  MIUR
(Cofin2001). S.C acknowledges financial support by MIUR (Cofin2002).
\end{acknowledgements}

\end{document}